\documentclass[12pt]{article}
\usepackage[polish,english]{babel}
\usepackage[latin1]{inputenc}
\usepackage{times}
\usepackage[T1]{fontenc}
\usepackage{epsfig}

\begin{document}

\title{Perturbations of the signum-Gordon model  }

\author{P. Klimas $\;$\\$\;\;$ \\  Institute of Physics,
Jagiellonian University, \\ Reymonta 4, 30-059 Cracow, Poland}

\date{$\;$}

\maketitle

\begin{abstract}
We investigate a perturbation of a scalar field model (called here
the signum-Gordon model) with the potential $V(f)=|f|$. The
perturbation generalizes the signum-Gordon model to the
signum-Klein-Gordon model i.e. to the case
$V(f)=|f|-\frac{1}{2}\lambda f^2$, where $\lambda$ is a small
parameter. Such a generalization breaks the scaling symmetry of
the signum-Gordon model. In this paper we concentrate on solutions
for self-similar initial data. Such data are particulary useful
for identification of the effects caused by the term that breaks
the scaling symmetry. We have found that the behaviour of the
solutions is quite interesting - they escape and return
periodically to the self-similar initial data.
\end{abstract}

\vspace*{2cm} \noindent PACS: 05.45.-a, 03.50.Kk, 11.10.Lm \\
\noindent Preprint TPJU - 15/2006

\pagebreak

\section{ Introduction}

The present paper refers to recently investigated scalar field
models with V-shaped potential \cite{AKTActa36}-\cite{Klimas}.
Such potentials have a common feature - left- and right hand
derivatives are different from zero at the minimum. Mentioned
models have a well-justified physical origin despite the fact that
they seem to be a little bit exotic from a mathematical viewpoint.
Moreover, some physical systems described by scalar field models
with V-shaped potentials are easy to built (e.g. chain of
pendulums impacting on a rectilinear bar). Unfortunately, such
models have a very unpleasant mathematical feature - a typical
solution consists of many (sometimes infinitely many) partial
solutions. The partial solutions are matched at some points. The
matching procedure is mostly onerous. This is probably the reason
why literature is poor in results for field theoretic models with
V-shaped potential.

It turns out that the behaviour of the field close to minimum
strongly depends on a 'shape' of the potential \cite{Arodz02}. In
particular, for V-shaped potential a field approaches exactly its
vacuum value at finite distance (a parabolic approach). This fact
has a profound significance - kinks have no exponential tails!
Such kinks are called {\it compactons} because their supports are
compact. The compactons considered in our models are topological,
so they are qualitatively different from e.g. well-known
compactons in the modified KdV model \cite{KdV1}-\cite{KdV2}.
Recently, the topological compactons have been also obtained in
models with nonstandard kinetic terms as so-called $k$-defects
\cite{ASW}. It is important to notice that there are other
(nontopological) compact solutions in the s-G model
\cite{oscylony}. Because of their properties they are called {\it
oscillons}.

The second characteristic property of the mentioned models is a
scaling symmetry, see \cite{AKTPR}. This symmetry means that if a
function $f(x,t)$ is a solution of a field equation, then new
function defined as $f_{\nu}(x,t)=\nu^2f(x/\nu,t/\nu)$ is a
solution as well. $\nu$ is here a positive constant. A presence of
the scaling symmetry in the model suggests existence of solutions
that are invariant with respect to the scaling transformation
(so-called self-similar solutions). Such solutions have been
obtained in the s-G model. A complete list of solutions for the
self-similar initial data is presented in \cite{AKTsignum}. For
models with the potential $V(f)=af\Theta(-f)+bf\Theta(f)$, where
$a$, $b$ are constant parameters and $\Theta$ is the well-known
Heaviside step function, the scaling symmetry is exact whereas for
most models with V-shaped potential the symmetry is only
approximated.  Note that the signum-Gordon model (s-G) can be
obtained as a particular case, i.e. by setting $a=-1$ and $b=1$. In
a group of models with symmetric V-shaped potential, the s-G model
is the simplest one. In this paper we study just symmetric
potentials.

The aim of the present work is the analysis of the perturbed s-G
model, where for simplicity reasons the specific perturbation is
chosen in the simplest, nontrivial form. Namely, we add the
quadratic term. In spite of its simplicity such a generalization
of the s-G model allows to face several important problems. The
first one is breaking of the scaling symmetry. Among the physical
systems there are fewer of them with an exact scaling symmetry.
There are always fluctuations in a typical physical system that
interacts with its environment. The fluctuations modify an
effective potential and break the exact scaling symmetry. In this
physical context, it is clear that investigation of the perturbed
field theoretic models with V-shaped potentials is an important
issue. It allows for better understanding of dynamics of
compactons in the systems with the broken scaling symmetry. In our
paper we analyse the perturbed potential
$V(f)=|f|-\frac{1}{2}\lambda f^2$, where $\lambda$ is a small
parameter i.e. $|\lambda|\ll 1$. The second important problem,
which is in general very difficult for systems with
non-differentiable potentials, is a stability analysis of
solutions. Our investigations are some kind of structural
stability analysis. Such analysis is important for compact kinks
as well as compact oscillons.

However, the models with V-shaped potentials are interesiting from the mathematical point of view, they have also some properties that allow to think about possible applications to condense matter physics and cosmology as well. In the cosmological context, the most interesiting seems to be the fact that for models with potentials sharp at its minima the terms that come from a gradient of the potential dominate the field dynamics close to the minimum. For instance, in the s-G model the term $\frac{dV}{df}=\hbox{sign}f$ remains finite arbitrary close to the minimum. This is in total oposition to the behaviour of e.g. $\phi^4$ theory, where gradient of the potential vanishes close to the minimum. Because of this, small perturbations propagate easily within the topological compactons or other nontopological field configurations like, e.g. mentioned oscillons, whereas outside of them the propagation encounters on resistance. Moreover, an absence of linear perturbations around the V-shaped minimum is a basic feature of our models. It entails automatically that the linear perturbations can propagate only at a defect background. This effect, characteriscic for models with V-shaped potentials, is similar to behaviour of $k$-fields that play a prominent role in cosmology (see \cite{AGSW}). 

In our calculation we concentrate on differences between solutions
in the s-G model and solutions in the signum-Klein-Gordon (s-K-G)
model (the perturbed model). Applying the same initial data for
solutions in both these models we can analyse the differences
between their solutions as a pure effect caused by the term
$\frac{1}{2}\lambda f^2$. In the case when initial data are
self-similar (parabolic) the solutions in the s-G model have
especially simple form. We apply the self-similar data just for
this reason.

Our paper is organized as follows. In the section 2 we
introduce the signum-Klein-Gordon (s-K-G) model and give a general
method of calculation of partial solutions that can be obtained
directly from the self-similar initial data. Unfortunately, they
are insufficient to construct a solution valid for each $x$ and
$t>t_0$, where $t_0$ is an initial moment. Section 3 is
devoted to a study of a solution for a specific self-similar
initial data. Focusing on a specific initial data enables us to
calculate all partial solutions that (when matched together) cover
the whole range of variable $x$. In the last section we summarize
our results and emphasize effects that stem from the term
$\frac{1}{2}\lambda f^2$.

\section{Initial problem for the generalized model}

\subsection{The signum-Klein-Gordon model and its partial solutions}

The s-K-G model for the scalar field $f(x,t)$ in 1+1 dimensions
has the Lagrangian
\begin{eqnarray}\label{lagranzjan}
L=\frac{1}{2}(\partial_t f)^2-\frac{1}{2}(\partial_x f)^2-V(f),
\end{eqnarray}
where the potential $V(f)$ is given by the formula
\begin{eqnarray}\label{potencjal}
V(f)=|f|-\frac{1}{2}\lambda f^2.
\end{eqnarray}
Euler-Lagrange equation that corresponds to Lagrangian
(\ref{lagranzjan}) takes the following form
\begin{eqnarray}\label{eqnonsamf}
(\partial^2_t-\partial^2_x)f+\hbox{sign}f-\lambda f=0.
\end{eqnarray}
The sign of parameter $\lambda$ has a crucial meaning for the
behaviour of the field $f(x,t)$. The potentials $V(f)$ for
negative and positive values of parameter $\lambda$ are
qualitatively different \cite{Klimas}. The case $\lambda = 0$
gives the s-G model which has been discussed in our previous
papers, (see e.g. \cite{AKTActa36}, \cite{AKTPR} and
\cite{AKTsignum}). In this paper we are interested in the case
$\lambda <0$, because for $\lambda>0$ the potential $V(f)$ is not
bounded from below. Nevertheless, a perturbative method presented
in the following subsection involves both cases of sign $\lambda$. In
order to distinguish between different kind of solutions, we use
symbol $f$ for solutions in the s-K-G model ($\lambda\neq0$) and
symbol $\phi$ for solutions in the s-G model. The partial
self-similar solutions are given by the formula
\begin{eqnarray}\label{rozwsam2}
\phi_k(x,t)=\frac{(-1)^k}{2}\frac{(x-v_{k-1}t)(x-v_{k}t)}{v_{k-1}v_{k}-1},
\end{eqnarray}
where $x\in [v_{k-1}t, v_kt]$ and $k=1,2,\ldots$. The parameters
$v_k$ are velocities of zeros of polynomials. They are determined
from matching conditions. For more details see \cite{AKTsignum}.
The partial solutions obey the relation
$\hbox{sign}\phi_k=(-1)^{k+1}$. By analogy, we define partial
solutions in the model with $\lambda\neq 0$. They obey the
equation
\begin{eqnarray}\label{eqnonsam}
(\partial^2_t-\partial^2_x)f_k(x,t)-(-1)^k-\lambda f_k(x,t)=0,
\end{eqnarray}
where $\hbox{sign}f_k=(-1)^{k+1}$. For $|\lambda|\ll 1$ the
potential $V(f)=|f|-\frac{1}{2}\lambda f^2$ can be interpreted as
a perturbed potential $V(\phi)=|\phi|$. In this case we say that
the exact scaling symmetry is violated or that the generalized
model has an approximate scaling symmetry when $|f|\gg \lambda
f^2$.

\subsection{Self-similar initial data and partial solutions}

This paper is devoted to investigation which are the effects caused
by the term $\lambda f$ in the s-K-G equation. It can be achieved
by comparison two solutions for the same initial data: the first
one that is a solution in the s-G model and the second one that
comes from the s-K-G model. The differences between them are a
direct consequence of the term that breaks the scaling symmetry.
From practical reasons we investigate some characteristic points
of solutions, i.e. trajectories of its zeros.

It turns out that explicit formulae for the solutions are not
always available - this problem strongly depends on initial data.
It has been shown, see \cite{AKTsignum}, that solutions in the s-G
model for the self-similar (parabolic) initial data are given by
explicit formulae. For this reason, the self-similar initial data
are more useful for our purposes than other, more general, initial
data. In fact, any self-similar solution $\phi$ at the moment
$t=t_0$ is suitable for our purposes and can be applied as an
initial data. Therefore, we assume following initial data for the
partial solutions
\begin{eqnarray}\label{warpocz1}
f_k(x,t_0)=\phi_k(x,t_0), \qquad \partial_t
f_k(x,t)|_{t=t_0}=\partial_t \phi_k(x,t)|_{t=t_0}.
\end{eqnarray}

\subsection{ The perturbative method}

The method presented in the current subsection allows us to obtain
the partial solutions $f_k(x,t)$ directly from the initial data
(\ref{warpocz1}). We call them {\it the partial solutions of the
first kind}. It turns out that such partial solutions are
insufficient. A complete solution $f(x,t)$ is consisted of some
other partial solutions as well. This inconvenience appears also
for, e.g. the s-G equation in the case when initial data have the
form of piecewise smooth functions matched up at some points. In
most cases, such matching points are origins of new partial
solutions. At the initial moment $t=t_0$ mentioned partial
solutions are shrunk to single points but for $t>t_0$ their
supports expand ({\it the partial solutions of the second kind}).
The s-K-G model has analogical partial solutions (the first and
the second kind) - we discuss their properties in the further part
of this paper. The partial solutions discussed in this paragraph
are the first kind ones. They cover whole range of axis $x$ at
$t=t_0$ (they obey (\ref{warpocz1}) where $\phi_k$ have this
property) and their supports shrink for $t>t_0$.

Let us assume that the solutions $f_k(x,t)$, which depend on
parameter $\lambda$, are represented in the form of a power series
\begin{eqnarray}\label{szereg}
f_k(x,t)=\sum_{n=0}^{\infty}f_{kn}(x,t)\lambda^n,
\end{eqnarray}
where $\lambda \ll 1$. After plugging series (\ref{szereg}) into
equation (\ref{eqnonsam}) we get a set of equations
\begin{eqnarray}\label{rownania}
\left\{
\begin{array}{ll}
(\partial^2_t-\partial^2_x)f_{k0}(x,t)=(-1)^k & n=0,\\
(\partial^2_t-\partial^2_x)f_{kn}(x,t)=f_{kn-1}(x,t) &
n=1,2,3,\ldots.
\end{array} \right.
\end{eqnarray}
Each of them takes the form of wave equation with a source. We can
integrate them using new variables $\xi=\frac{1}{2}(x+t)$,
$\eta=\frac{1}{2}(x-t)$. The result of integration in the original
variables reads
\begin{eqnarray}\label{fk0}
f_{k0}(x,t)=F_0(x+t)+G_0(x-t)-\frac{(-1)^k}{4}(x^2-t^2),
\end{eqnarray}
{\setlength\arraycolsep{2pt}
\begin{eqnarray}\label{fkn}
f_{kn}(x,t)&=&F_n(x+t)+G_n(x-t)-\int_0^{\frac{x+t}{2}}\hbox{d}\alpha\int_0^{\frac{x-t}{2}}
\hbox{d}\beta\; f_{k n-1}(\alpha+\beta,\alpha-\beta),\nonumber \\
\end{eqnarray}}
where $F(x+t)$ and $G(x-t)$ are arbitrary functions. They can be
calculated from the following initial conditions for the partial
solutions:
\begin{eqnarray}\label{warpocz2}
f_{k0}(x,t_0)=\phi_k(x,t_0), \qquad \partial_t
f_{k0}(x,t)|_{t=t_0}=\partial_t \phi_k(x,t)|_{t=t_0},
\end{eqnarray}
\begin{eqnarray}\label{warpocz3}
f_{kn}(x,t_0)=0, \qquad \partial_t f_{kn}(x,t)|_{t=t_0}=0.
\end{eqnarray}
Conditions (\ref{warpocz2}) and (\ref{warpocz3}) stem from the
initial conditions (\ref{warpocz1}). The solution (\ref{fk0})
obeys the first of the equations (\ref{rownania}) (i.e. the s-G
equation) and the self-similar initial data (\ref{warpocz2}), so
it coincides with $\phi_k$,
\begin{eqnarray}\label{rozw0}
f_{k0}(x,t)=\phi_k(x,t).
\end{eqnarray}
A direct calculation confirms this result. It means that the
self-similar solutions (\ref{rozw0}) are zero-order approximation
for the solutions (\ref{szereg}). In order to get higher-order
approximations we have to find the functions $F_n(x+t)$ and
$G_n(x-t)$. Differentiating first of equations
(\ref{warpocz3}) with respect to $x$, combining with the second
one and shifting arguments, we obtain equations
\begin{eqnarray}\label{rownaniaD}
D_{+}f_{kn}(x,t)\left.\right|_{x=s-t_0,\;t=t_0}=0,\qquad
D_{-}f_{kn}(x,t)\left.\right|_{x=w+t_0,\;t=t_0}=0,
\end{eqnarray}
where $D_{\pm}\equiv \frac{1}{2}(\partial_x\pm
\partial_t)$. Equations (\ref{rownaniaD}) can be rewritten in the
form
\begin{eqnarray}\label{F'n}
F'_n(s)=\left.D_+\int_0^{\frac{x+t}{2}}\hbox{d}\alpha\int_0^{\frac{x-t}{2}}
\hbox{d}\beta\; f_{k
n-1}(\alpha+\beta,\alpha-\beta)\right|_{x=s-t_0,\;t=t_0},
\end{eqnarray}
\begin{eqnarray}\label{G'n}
G'_n(w)=\left.D_-\int_0^{\frac{x+t}{2}}\hbox{d}\alpha\int_0^{\frac{x-t}{2}}
\hbox{d}\beta\; f_{k
n-1}(\alpha+\beta,\alpha-\beta)\right|_{x=w+t_0,\;t=t_0},
\end{eqnarray}
where formula (\ref{fkn}) has been applied. The sum of constants
that comes from integration of the expressions (\ref{F'n}) and
(\ref{G'n}) is fixed by the first of conditions (\ref{warpocz3}).

In the first step we calculate the function $f_{k1}(x,t)$ from
formula (\ref{fkn}). Then we can continue the procedure in order
to obtain $f_{k2}(x,t)$. In principle, this procedure can be
repeated infinitely many times giving expressions for all
functions $f_{kn}(x,t)$. In fact, we are able to obtain only few
functions $f_{kn}(x,t)$ because the calculations became quickly
too complicated. Fortunately, functions $f_{kn}(x,t)$ obtained for
several, the lowest values of $n$ enable us to guess a general
formula for arbitrary $n$. This formula has the form
\begin{eqnarray}\label{rozwn}
f_{kn}(x,t)=(-1)^k\frac{(t-t_0)^{2n}}{(2n+2)!}\frac{A_nx^2+B_nx+C_n}{v_{k-1}v_{k}-1},
\end{eqnarray}
where
\[
A_n=(2n+1)(n+1),
\]
\[
B_n=-(v_{k-1}+v_{k})(n+1)(t+2nt_0),
\]
\[
C_n=(n+v_{k-1}v_{k})t^2+n(1+(2n+1)v_{k-1}v_{k})t_0^2.
\]
One can check that formula (\ref{rozwn}), which was originally
found for $n=1,2,\ldots$, is also true for $n=0$ - in this case it
gives (\ref{rozw0}).

It turns out, and this is a big surprise, that the series
(\ref{szereg}) can be summed up giving as a result
\begin{eqnarray}\label{rozw_lambda>0}
f_k(x,t)&=&\frac{(-1)^k}{2(v_{k-1}v_{k}-1)}\left[
M\cosh{(\rho(t-t_0))}+N\frac{\sinh{(\rho(t-t_0))}}{\rho}\right]+
\nonumber\\
&+& \frac{(-1)^k}{\rho^2}[\cosh{(\rho(t-t_0))}-1],
\end{eqnarray}
for $\lambda\equiv\rho^2>0$, and
\begin{eqnarray}\label{rozw_lambda<0}
f_k(x,t)&=&\frac{(-1)^k}{2(v_{k-1}v_{k}-1)}\left[
M\cos{(\sigma(t-t_0))}+N\frac{\sin{(\sigma(t-t_0))}}{\sigma}\right]+
\nonumber\\
&+& \frac{(-1)^k}{\sigma^2}[1-\cos{(\sigma(t-t_0))}],
\end{eqnarray}
for $\lambda\equiv-\sigma^2<0$. The coefficients $M$ and $N$ read
\[
M\equiv(x-v_{k-1}t_0)(x-v_{k}t_0),
\]
\[
N\equiv t-t_0+2v_{k-1}v_{k}t_0-(v_{k-1}+v_{k})x.
\]
The partial solutions $f_k$ satisfy the relation
$\hbox{sign}f_k=(-1)^{k+1}$.

It is worth emphasizing that the partial solutions $f_k$ in the
model with the explicitly broken scaling symmetry, still have a
quadratic dependence on variable $x$. What changes is the time
dependence. This is the first important result obtained with the
help of the perturbative method.

Knowing this result we can, of course, propose {\it a posteriori}
a proper Ansatz
\begin{eqnarray}\label{ansatz}
f_k(x,t)=a(t)x^2+b(t)x+c(t).
\end{eqnarray}
Plugging the Ansatz (\ref{ansatz}) into eq. (\ref{eqnonsam}) we
get the set of ordinary differential equations for the
coefficients $a(t)$, $b(t)$ and $c(t)$
\[
\frac{d^2a}{dt^2}-\lambda a=0,\qquad \frac{d^2b}{dt^2}-\lambda
b=0, \qquad \frac{d^2c}{dt^2}-\lambda c=2a+(-1)^k.
\]
The constants that come from integration of these equations are
fixed by the condition (\ref{warpocz1}). Solving these equations
we recover formulae (\ref{rozw_lambda>0}) and
(\ref{rozw_lambda<0}) depending on sign of the parameter
$\lambda$.

\section{The solution for a specific initial data}

In this section we present a solution in the s-K-G model for a
specific self-similar initial data. We concentrate on the case
when the perturbative parameter $\lambda$ is small $|\lambda|\ll
1$ and negative $\lambda = -\sigma^2$. The main effort is focused
on the partial solutions of the second kind. They can be obtained
as the solutions of boundary problem. The boundary conditions are
given by the partial solutions of the first kind at points of
contact of their supports and supports of the partial solutions of
the second kind. The most serious obstacle in the s-K-G model is
that we do not know a general formula for the partial solutions.
For instance, in the s-G model such formula consists of two
arbitrary functions and terms $\pm t^2/2$, $\pm x^2/2$ or their
combinations. For this reason, we restrict our study to a specific
self-similar initial data and construct an approximated solution.
Nevertheless, it turns out that investigation of such a specified
case gives valuable information as well. One of the most
interesting results presented in the current section is a
discovery of periodicity in time for such solution in the s-K-G
model.

\subsection{The positive partial solution $f_+(x,t)$}

\begin{figure}[h!]
\begin{center}
\includegraphics[width=0.7\textwidth]{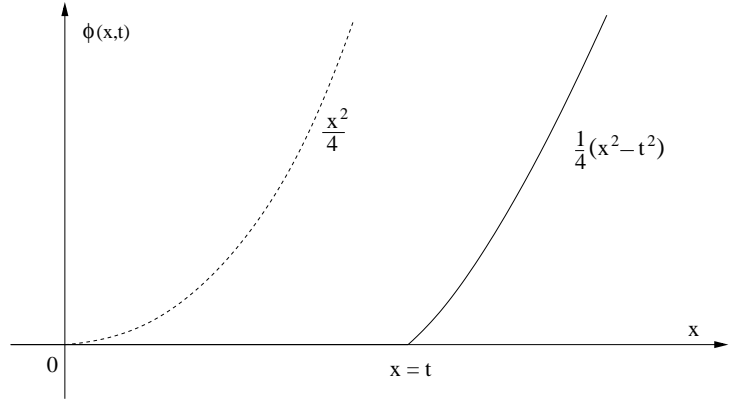}
\caption{The self-similar solution for initial data
(\ref{warpocz}). The dashed line represents an initial
configuration of the field $\phi$.}\label{sssolution}
\end{center}
\end{figure}

Among the self-similar initial data, the simplest one reads
\begin{eqnarray}\label{warpocz}
\phi(x,0)=\frac{1}{4}x^2\Theta(x),\qquad \partial_t
\phi(x,t)|_{t=0}=0,
\end{eqnarray}
where $\Theta (x)$ is the Heaviside step function. The
self-similar solution $\phi(x,t)$ in the s-G model for data
(\ref{warpocz}) is consisted of two partial solutions matched up
at the light cone $x=t$, see also \cite{AKTsignum}. The solution
$\phi(x,t)$ has a very simple form
\begin{eqnarray}\label{ss}
\phi(x,t)=\frac{1}{4}(x^2-t^2)\Theta(x-t).
\end{eqnarray}
This formula corresponds to (\ref{rozwsam2}) for $k=1$, $v_0=-1$
and $v_1=1$. The snapshot of the solution $\phi(x,t)$ is presented
in Fig.\ref{sssolution}.

In the further part of this paper we concentrate on the solution
$f(x,t)$ in the s-K-G model. The positive partial solution
$f_+(x,t)$ obeys the equation
\begin{eqnarray}\label{eq}
(\partial^2_t-\partial^2_x)f_++\sigma^2 f_++1=0,
\end{eqnarray}
and the initial conditions
\begin{eqnarray}
f_+(x,0)=\frac{1}{4}x^2,\qquad \partial_t
f_+(x,t)|_{t=0}=0.\nonumber
\end{eqnarray}
It takes the form
\begin{eqnarray}\label{rozwnonsam3}
f_+(x,t)=\frac{1}{4}\cos(\sigma t)x^2+\frac{t}{4\sigma}\sin(\sigma
t)-\frac{1}{\sigma^2}(1-\cos(\sigma t)).
\end{eqnarray}
The partial solution (\ref{rozwnonsam3}) can be obtained directly
from (\ref{rozw_lambda<0}) for $k=1$ and $t_0 = 0$. The formula
(\ref{rozwnonsam3}) holds for $x\geq x_1(t)$. The trajectory of
zero $x_1(t)$,
\begin{eqnarray}\label{zero1}
x_1(t)=\frac{2}{\sigma}\sqrt{\frac{1}{\cos(\sigma t
)}-\frac{\sigma t}{4}\tan(\sigma t)-1}
\end{eqnarray}
is the solution of the equation $f_+(x_1,t)=0$. Except for the
point $t=0$, the function $x_1(t)$ obeys inequality $x_1(t)>t$,
what means that zero of $f_+(x,t)$ moves faster than its
counterpart ($x(t)=t$) in the s-G model. Moreover, we can see that
the velocity of zero $x_1(t)$ depends on variable $t$. Let us
remind that velocities $v_k$ of zeros of the self-similar
solutions are constant, see formula (\ref{rozwsam2}). It means
that the term $\lambda f$ in the eq. (\ref{eqnonsamf}) is
responsible for non monotonous expansion or contraction of the
supports of the partial solutions.
\begin{figure}[h!]
\begin{center}
\includegraphics[width=0.8\textwidth]{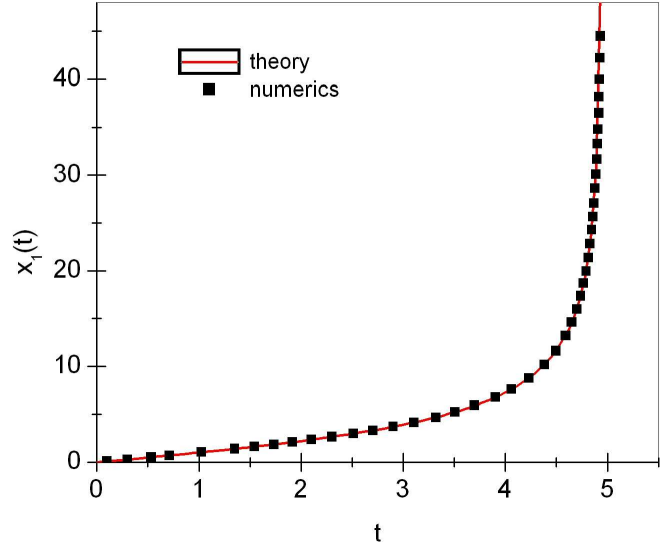}
\caption{The trajectory $x_1(t)$ for $\lambda=-0.1$.}\label{rysx0}
\end{center}
\end{figure}

A series expansion of the formula (\ref{zero1}) for small $t$
\begin{eqnarray}\label{x1_rozwiniecie}
x_1(t)=t+\frac{1}{4}t^3\sigma^2+\frac{103}{1440}t^5\sigma^4+O(t^7)
\end{eqnarray}
gives valuable information as well. We can see from
(\ref{x1_rozwiniecie}) that the zero $x_1(t)$ moves with the
acceleration $\ddot{x}_1(t)$ which is proportional to $\sigma^2$
provided that $t\ll 1$. This observation has a practical meaning
because it enables us to calculate the parameter $\lambda$ from
experimental data. The curve $x_1(t)$ is presented in the Fig.
\ref{rysx0}. We can see a very good agreement between the
analytical calculation and the numerical data. The function
(\ref{zero1}) goes to infinity for $t\rightarrow
t^*\equiv\frac{\pi}{2\sigma}$. It means that the solution
$f_+(x,t)$ is valid for $t<t^*$. In our numerical calculation
($\sigma^2=0.1$) the characteristic time $t^*\approx 4.9673$. The
leading behaviour of $x_1(t)$ close to $t^*$ is given by the first
term of the expansion
\begin{eqnarray}
x_1(t)=\frac{1}{2\sigma^{3/2}}\sqrt{\frac{16-2\pi}{t^*-t}}-
\frac{3}{\sigma^{1/2}}\sqrt{\frac{t^*-t}{16-2\pi}}+O(\sigma^{1/2}).\nonumber
\end{eqnarray}

\subsection{The negative partial solution $f_-(x,t)$ - some general remarks}

In this and two further paragraphes we present the partial
solution of the second kind $f_-(x,t)$. Our solution is only
approximated and holds for times not longer than $t\approx 2$.
Such a partial solution can be obtained as the solution of
boundary problem because at the initial time $t=0$ its support is
a single point and it is located at $x=0$. For later times $t>0$
the support expands to infinite size. This behaviour has been
observed in our numerical calculation and it is suggested by the
fact that $x_1(t)$ tends to infinity for $t\rightarrow
\frac{\pi}{2\sigma}$. The solution $f_-(x,t)$ has to obey the
following boundary conditions
\begin{eqnarray}\label{war_zszycia1}
f_-(x_1,t)=0,\qquad \partial_x f_-(x,t)|_{x=x_1}=\partial_x
f_+(x,t)|_{x=x_1},
\end{eqnarray}
\begin{eqnarray}\label{war_zszycia2}
f_-(x_0,t)=0,\qquad \partial_x f_-(x,t)|_{x=x_0}=0.
\end{eqnarray}
Conditions (\ref{war_zszycia1}) and (\ref{war_zszycia2}) are
derived from the field equation (\ref{eqnonsamf}) and they mean
that the partial solutions are matched so that the solution
$f(x,t)$ is smooth at $x_1(t)$ and $x_0(t)$. At $x_0(t)$ the
partial solution $f_-(x,t)$ is matched with the trivial partial
solution $f_0(x,t)=0$. Whereas the point $x_1(t)$ is given by the
formula (\ref{zero1}), the second zero of $f(x,t)$, i.e $x_0(t)$,
is not known yet. In Section 3.4 we show how to obtain an
approximated formula for $x_0(t)$.

\begin{figure}[h!]
\begin{center}
\includegraphics[width=0.65\textwidth]{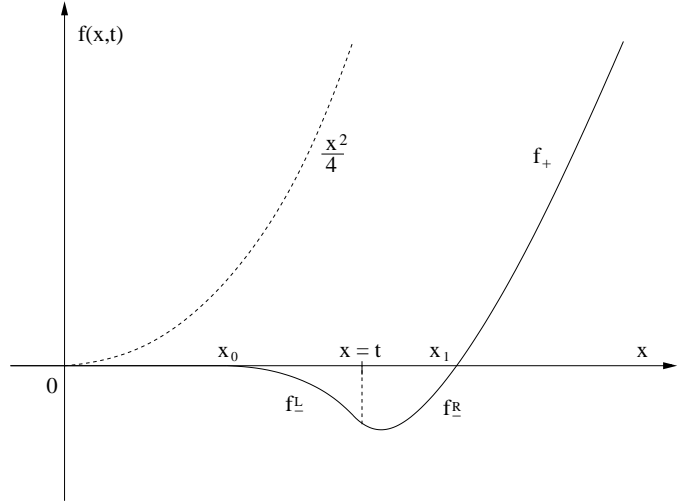}
\caption{The solution $f(x,t)$ for initial data (\ref{warpocz})
and times $t<\frac{\pi}{2\sigma}$ The dashed line represents an
initial configuration of the field $f$.}\label{nssolution}
\end{center}
\end{figure}

It turns out that the partial solution that obeys
(\ref{war_zszycia1}) does not obey (\ref{war_zszycia2}) and vice
versa. In order to get rid of this inconvenience we split the
solution $f_-(x,t)$ into two pieces $f_-^L(x,t)$ and $f_-^R(x,t)$.
They are matched at the light cone $x=t$. Such split is sufficient
to obtain $f_-(x,t)$ that obeys (\ref{war_zszycia1}) and
(\ref{war_zszycia2}) simultaneously. The solution $f(x,t)$
consists of the following partial solutions
\[
f(x,t) = \left\{
\begin{array}{ll}
0 &\hbox{for}\quad x\leq x_0(t),\\
f_{-}^L(x,t) &\hbox{for}\quad x_0(t)\leq x\leq t, \\
f_{-}^R(x,t) &\hbox{for}\quad t \leq x\leq x_1(t), \\
f_+(x,t) &\hbox{for}\quad x\geq  x_1(t),
\end{array} \right.
\]
where $t<\frac{\pi}{2\sigma}$. The snapshot of $f(x,t)$ is
depicted in Fig. \ref{nssolution} - compare it to the solution in
Fig. \ref{sssolution}.

\subsection{The partial solution $f_-^R(x,t)$}

It has been already mentioned at the begining of Section 3 that a
general formula for the partial solutions in the s-K-G model is
not known. This is the most serious obstacle in our
investigations. Therefore, we search for the approximated partial
solution $f_-(x,t)$. The approximation of eq. (\ref{eqnonsamf}) is
obtained by replacing the term $\sigma^2 f(x,t)$ by the term
$\sigma^2 \phi(x,t)$, what gives us
\begin{eqnarray}\label{eq_approx}
(\partial^2_t-\partial^2_x)f_-(x,t)+\sigma^2 \phi(x,t)-1=0.
\end{eqnarray}
Such a modification is valid only for small times. The partial
solution $f_-^R(x,t)$ of eq. (\ref{eq_approx}) (an approximate
solution of eq. (\ref{eqnonsamf}) for $\lambda=-\sigma^2$) at
$t\leq x\leq x_1(t)$ has the form
\begin{eqnarray}\label{f-R}
f_-^R(x,t)=F_{R}(x+t)+G_{R}(x-t)-\frac{1}{4}(x^2-t^2)+\frac{\sigma^2}{64}(x^2-t^2)^2.
\end{eqnarray}
In accordance with (\ref{war_zszycia1}), this solution is matched
to $f_+(x,t)$ at $x=x_1(t)$. It turns out that exact formulas for
$F_{R}(x+t)$ and $G_{R}(x-t)$ can not be achieved because we need
the inverse functions of $x_1(t)\pm t$, where $x_1(t)$ is given by
(\ref{zero1}). Nevertheless, we can expand the expressions
$x_1(t)\pm t$ in power series and then invert these series up to
an arbitrary term. This is why we concentrate on series expansions
of the partial solutions. In our further calculations we use the
perturbation parameter $\sigma$ as an expansion parameter. The
partial solutions represented by finite series (i.e. approximated
partial solutions) obey the matching conditions
(\ref{war_zszycia1}) and (\ref{war_zszycia2}) up to some range of
$\sigma$. In order to find this range we start from series
expansion of the formula (\ref{zero1}) for $\sigma \ll 1$, i.e.,
\begin{eqnarray}\label{x1_Rozwiniecie}
x_1(t)=t+\frac{1}{4}t^3\sigma^2+\frac{103}{1440}t^5\sigma^4+O(\sigma^6).
\end{eqnarray}
Note that the expansion (\ref{x1_Rozwiniecie}) has the same form
as the expansion (\ref{x1_rozwiniecie}) for small times $t$. A
leading term of the expression $x_1(t)-t$ is proportional to
$\sigma^2$. The powers of the expression $x_1(t)-t$ appear in
(\ref{war_zszycia1}) because the partial solution $f_-^R(x,t)$,
which is given by formula (\ref{f-R}), includes terms proportional
to $x-t$ and $\sigma^2(x-t)^2$. In order to take into
consideration contributions from all terms in (\ref{f-R}),
especially from $\sigma^2(x-t)^2$, we need terms proportional to
$\sigma^6$ at least. This is, naively, an accuracy of
$f_-^R(x,t)$. The real accuracy is lower. The direct calculations
allow us to obtain $f_-^R(x,t)$ only up to terms proportional to
$\sigma^4$ because solutions of (\ref{war_zszycia1}), i.e. $F'_R$
and $G'_R$ have such accuracy, see formulae below. The prime '
stands for derivatives with respect to whole arguments of $F_R$
and $G_R$.

In the first step we expand formulae (\ref{rozwnonsam3}),
(\ref{f-R}) and their derivatives with respect to variable $x$ at
$x=x_1(t)$, where $x_1(t)$ is given by the formula
(\ref{x1_Rozwiniecie}). The result has the following form

{\setlength\arraycolsep{2pt}
\begin{eqnarray}
f_+(x_1(t),t)&=&O(\sigma^6),\nonumber
\end{eqnarray}
\begin{eqnarray}
\partial_xf_+(x,t)|_{x=x_1}&=&\frac{1}{2}t-\frac{1}{8}t^3\sigma^2-\frac{17}{2880}t^5\sigma^4+O(\sigma^6),\nonumber
\end{eqnarray}
{\setlength\arraycolsep{2pt}
\begin{eqnarray}
f_-^R(x_1,t)&=&F_R(x_1+t)+G_R(x_1-t)-\frac{1}{8}t^4\sigma^2-\frac{37}{720}t^6\sigma^4+O(\sigma^6),\nonumber
\end{eqnarray}
\begin{eqnarray}
\partial_xf_-^R(x,t)|_{x=x_1}&=&F'_R(x_1+t)+G'_R(x_1-t)
-\frac{1}{2}t-\frac{1}{8}t^3\sigma^2-\frac{13}{2880}t^5\sigma^4+O(\sigma^6).\nonumber
\end{eqnarray}
Plugging three last formulae into conditions (\ref{war_zszycia1})
we obtain two equations that contain $F_R$, $G_R$, $F'_R$ and
$G'_R$. Then, we differentiate the first of these equations, i.e.
$f_-(x_1(t),t)=0$ with respect to variable $t$. In the next step
we solve obtained equations with respect to $F'_R$ and $G'_R$,
which gives
\begin{eqnarray}\label{FF'}
F'_R(x_1+t)&=&-\frac{1}{8}t^3\sigma^2-\frac{71}{2880}t^5\sigma^4+O(\sigma^6),
\end{eqnarray}
\begin{eqnarray}\label{GG'}
G'_R(x_1-t)&=&t+\frac{1}{8}t^3\sigma^2+\frac{67}{2880}t^5\sigma^4+O(\sigma^6).
\end{eqnarray}
Eq. (\ref{FF'}) can be integrated with the help of new variable
$s=x_1(t)+t$, where $x_1(t)$ is given by series
(\ref{x1_Rozwiniecie}). In the inverse series
\begin{eqnarray}\label{t(s)}
t(s)=\sum_{k=0}^Nb_ks^{2k+1}\sigma^{2k},
\end{eqnarray}
only terms up to $N=1$ are significant to ensure the given
accuracy. Coefficients $b_0$ and $b_1$ have the following
numerical values
\[
b_0=\frac{1}{2},\qquad b_1=-\frac{1}{64}.
\]
Consequently, the approximate formula for $F_R(s)$ takes the form
\begin{eqnarray}
F_R(s)=-\frac{1}{256}s^4\sigma^2+\frac{1}{8640}s^6\sigma^4+O(\sigma^6).
\end{eqnarray}
In the similar way we compute $G_R(w)$, where $w=x_1(t)-t$. The
inverse series is given by the formula
\begin{eqnarray}\label{t(w)}
t(w)=\sum_{k=0}^Nc_kw^{\frac{2k+1}{3}}\sigma^{\frac{2k-2}{3}}.
\end{eqnarray}
In this case we have to compute coefficient $c_k$ up to $N=7$. We
will not present here their numerical values. Function $G_R(w)$
takes the form
\begin{eqnarray}
G_R(w)=\sum_{k=0}^7g_kw^{\frac{2k+4}{3}}\sigma^{\frac{2k-2}{3}},\nonumber
\end{eqnarray}
where the coefficients $g_k$ have the following approximated
numerical values:
\begin{eqnarray}
g_0=1.1900,\qquad g_1=0.0593,\qquad g_2=-0.0299,\qquad
g_3=-0.0429,\nonumber
\end{eqnarray}
\begin{eqnarray}
g_4=0.0037,\qquad g_5=-0.0134,\qquad g_6=0.0022,\qquad
g_7=0.0002.\nonumber
\end{eqnarray}

It is important to note that the term $k=0$ in $G_R(w)$ has a
singular dependence on $\sigma$, i.e. it is proportional to
$\sigma^{-2/3}$. Such behaviour is caused by the fact that term
proportional to $\sigma^0$, linear in $t$, is cancelled in the
definition of the variable $w$ which is proportional to $\sigma^2
t^3$ for $t\ll 1 $. In fact, the singular term $\sigma^{-2/3}$
appears already in the series (\ref{t(w)}).

\subsection{The partial solution $f_-^L(x,t)$}

For $x\leq t$ the self-similar solution (\ref{ss}) is equal to zero,
therefore eq. (\ref{eq_approx}) takes the form
\begin{eqnarray}\label{eqfL}
(\partial^2_t-\partial^2_x)f_-^L(x,t)-1=0.
\end{eqnarray}
The general solution of (\ref{eqfL}) is of the form
\begin{eqnarray}\label{f_-^L}
f_-^L(x,t)=F_{L}(x+t)+G_{L}(x-t)-\frac{1}{4}(x^2-t^2).
\end{eqnarray}
The arbitrary functions $F_L$, $G_L$ and unknown function $x_0(t)$
can be obtained after imposing the following matching conditions:
\begin{eqnarray}\label{secondconditions}
f_-^L(t,t)=f_-^R(t,t),\qquad
f_-^L(x_0,t)=0,\qquad\partial_xf_-^L(x,t)|_{x=x_0}=0.
\end{eqnarray}
The first of conditions (\ref{secondconditions}) gives equality of
values of the partial solutions $f_-^R$ and $f_-^L$ at the light
cone $x=t$. We do not require equality of spatial derivatives of
these partial solutions but it turns out that equality of values
entails equality of derivatives as well. The matching condition at
$x=t$ gives an equality $F_L(2t)+G_L(0)=F_R(2t)$, which allows us
to obtain the function $F_L$. Without loosing of generality we can
fix $G_L(0)=0$, because $G_L(0)$ is cancelled in the combination
$F_{L}(x+t)+G_{L}(x-t)$. Last two matching conditions
(\ref{secondconditions}) allow us to obtain derivatives of
functions $F_L$ and $G_L$. Differentiating the second condition in
(\ref{secondconditions}) with respect to variable $t$ and
combining with the third one we obtain
\begin{eqnarray}\label{rownanianafL}
D_+f_-^{L}(x,t)\left.\right|_{x=s-t(s),\;t=t(s)}=0,\qquad
D_-f_-^{L}(x,t)\left.\right|_{x=w+t(w),\;t=t(w)}=0,
\end{eqnarray}
where $s=x_0(t)+t$, $w=x_0(t)-t$ and
$D_\pm\equiv\frac{1}{2}(\partial_x\pm\partial_t)$. The solutions
of (\ref{rownanianafL}) take the form
\begin{eqnarray}\label{F'(s)G'(w)}
F'_{L}(s)=\frac{1}{4}(s-2t(s)),\qquad
G'_{L}(w)=\frac{1}{4}(w+2t(w)).
\end{eqnarray}

In order to obtain $x_0(t)$, we use the equality
$F'_L(s)=F'_R(s)$, what gives the equation
\[
\frac{\sigma^4}{45}s^5-\frac{\sigma^2}{2}s^3-8s+16t=0.
\]
A solution of this equation $s(t)$ can be obtained in the series
form. Finally, it gives $x_0(t)$ in the form
\begin{eqnarray}\label{x0(t)}
x_0(t)=t-\frac{1}{2}t^3\sigma^2+\frac{167}{360}t^5\sigma^4+O(\sigma^6).
\end{eqnarray}
A leading term of deceleration of zero $x_0$ is proportional to
$\sigma^2$ for $t\ll 1$.
\begin{figure}[h!]
\begin{center}
\includegraphics[width=0.8\textwidth]{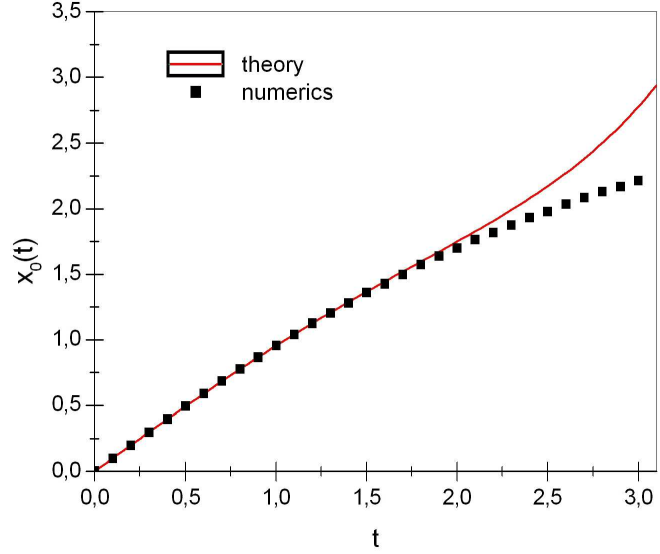}
\caption{Trajectory of $x_0(t)$.}\label{x0}
\end{center}
\end{figure}
Fig. \ref{x0} depicts the trajectory of $x_0(t)$. The analytical
curve which is given by the first three terms in formula
(\ref{x0(t)}) is a good approximation for a numerical trajectory
if $t$ is not greater than $t\approx 2$. It gives the limitation
for validity of the approximated partial solution $f_-(x,t)$.

The explicit formula for $x_0(t)$ enables us to obtain the
function $G_L$ and the solution $f_-^L(x,t)$. Function $x_0(t)$,
given by (\ref{x0(t)}), is known up to $O(\sigma^4)$ (including
this term) so $t(w)$ can be obtained up to $O(\sigma^0)$. The
second formula in (\ref{F'(s)G'(w)}) gives
\begin{eqnarray}\label{GL}
G_L(w)=-\frac{3}{8}\left(\frac{2}{\sigma^2}\right)^{1/3}w^{4/3}-\frac{4}{135}w^2+O(\sigma^{2/3}).
\end{eqnarray}
Finally, we obtain the partial solution $f_-^{L}(x,t)$ of the form
\begin{eqnarray}\label{fL}
f_-^{L}(x,t)=&-&\frac{1}{540}(151x-119t)(x-t)-\frac{\sigma^2}{256}(x+t)^2+\frac{\sigma^4}{8640}(x+t)^4-\nonumber\\
&-&\frac{3}{8}\left(\frac{2}{\sigma^2}\right)^{1/3}(x-t)^{4/3}.
\end{eqnarray}

\begin{figure}[h!]
\begin{center}
\includegraphics[width=0.65\textwidth]{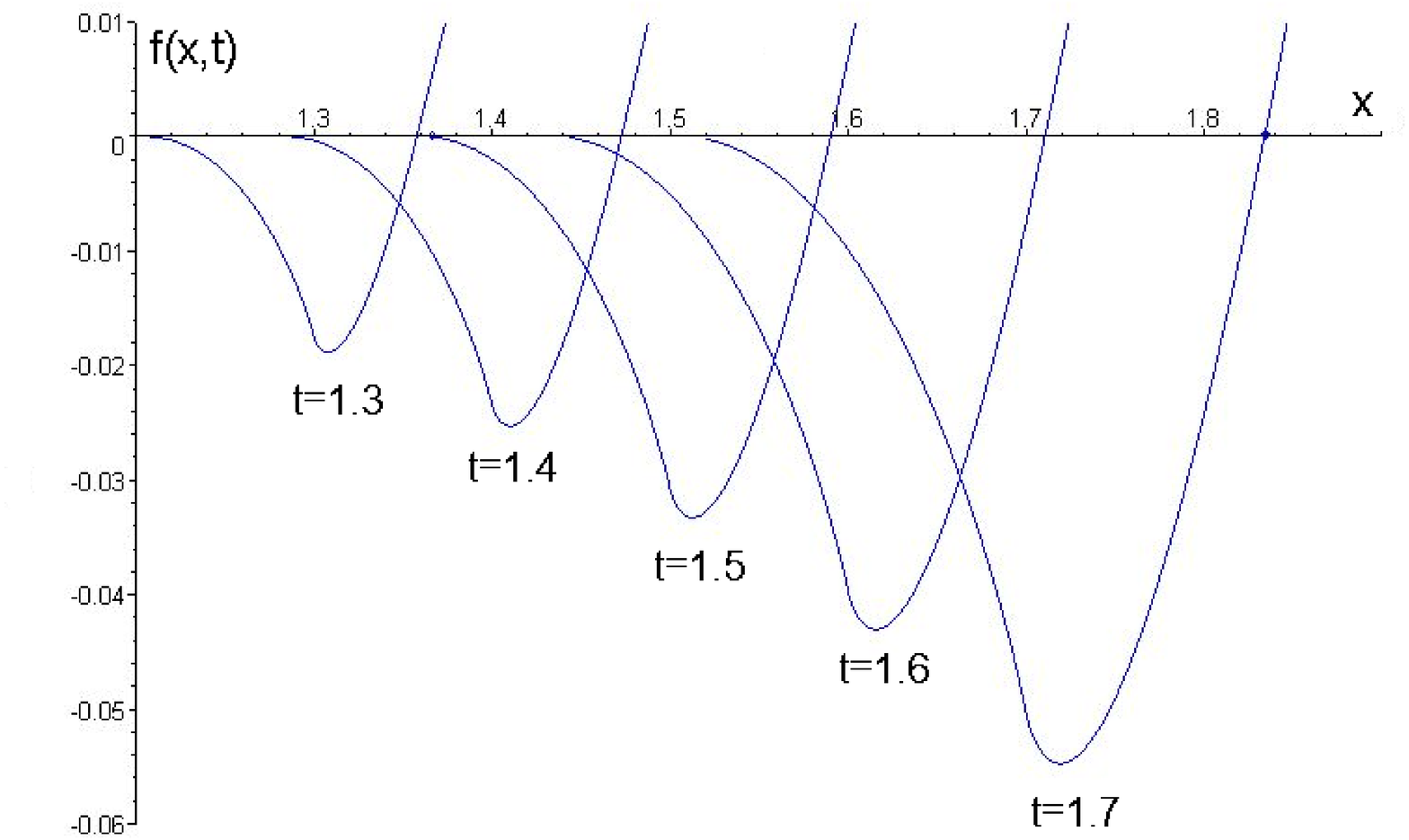}
\caption{Evolution of approximated analytical solution $f(x,t)$
for $\sigma^2=0.1$.}\label{absf}
\end{center}
\end{figure}

\begin{figure}[h!]
\begin{center}
\includegraphics[width=0.8\textwidth]{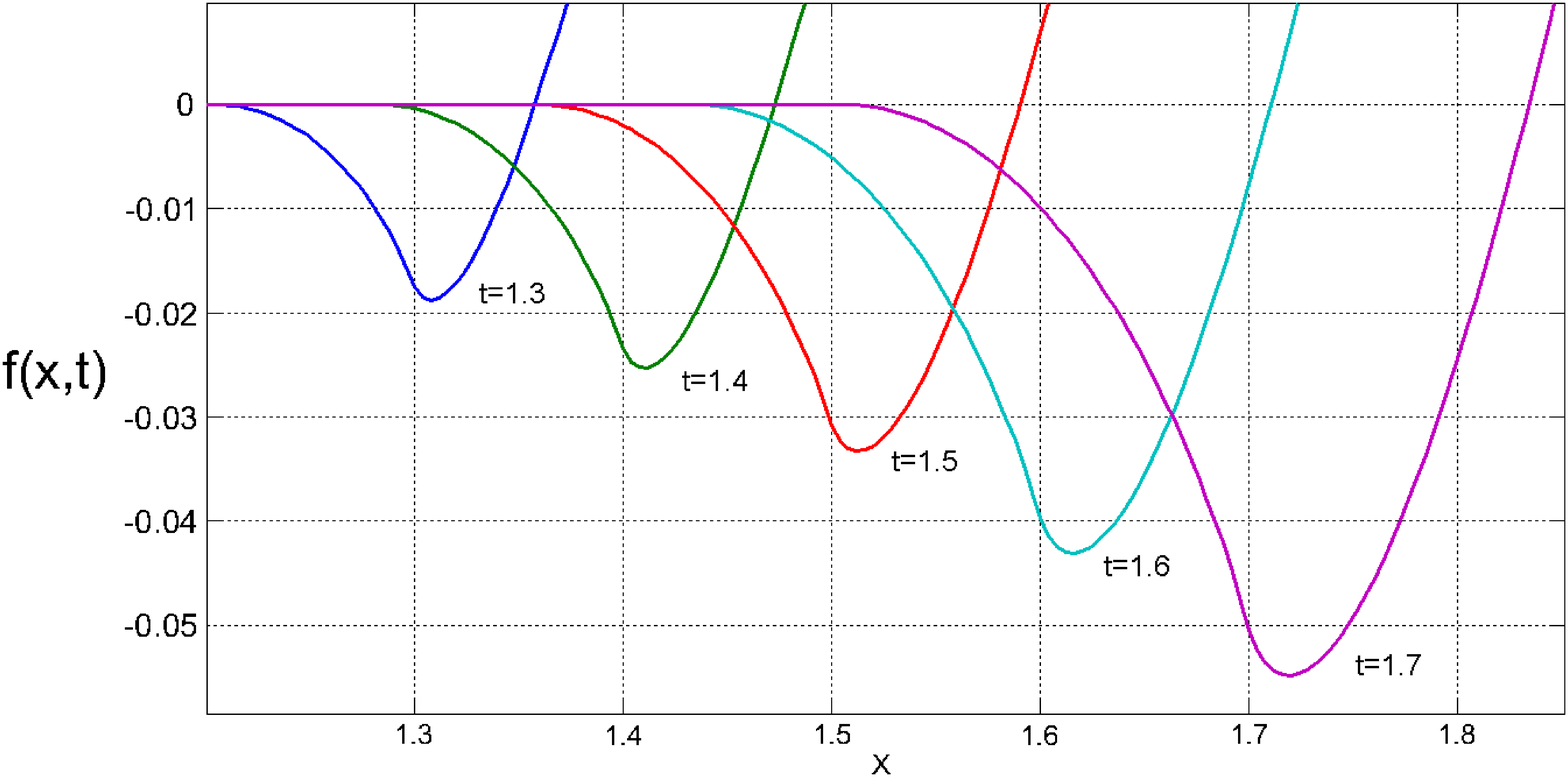}
\caption{Numerical solution $f(x,t)$ for
$\sigma^2=0.1$.}\label{numabsf}
\end{center}
\end{figure}

We note that the singular term $\sim\sigma^{-2/3}$ is present in
the formulae (\ref{GL}) and (\ref{fL}). There is nothing
unexpected in this fact because for small times the function
$x_0(t)-t$ has similar behaviour to the function $x_1(t)-t$ from
the previous paragraph. Some snapshots of the solution $f(x,t)$
for first stage of evolution are presented in Figs. \ref{absf} and
\ref{numabsf}. There is a very good agreement between numerical
and analytical solutions untill $t\approx 2$.

\subsection{Behaviour for later times $t>t^*$}

In the current section we study a numerical solution for the s-K-G
model with $\lambda = -0.1$. The solution at $t=0$ obeys following
conditions
\begin{eqnarray}\label{dane}
f(x,0)=\frac{1}{4}x^2\Theta(x),\qquad \partial_t
f(x,t)|_{t=0}=0.\nonumber
\end{eqnarray}
We focus on times $t>t^*$, where $t^*\approx 4.9673$. The
trajectories of zeros of $f(x,t)$ up to $t=25$ are depicted in
Fig. \ref{duze czasy}.
\begin{figure}[h!]
\begin{center}
\includegraphics[width=0.8\textwidth]{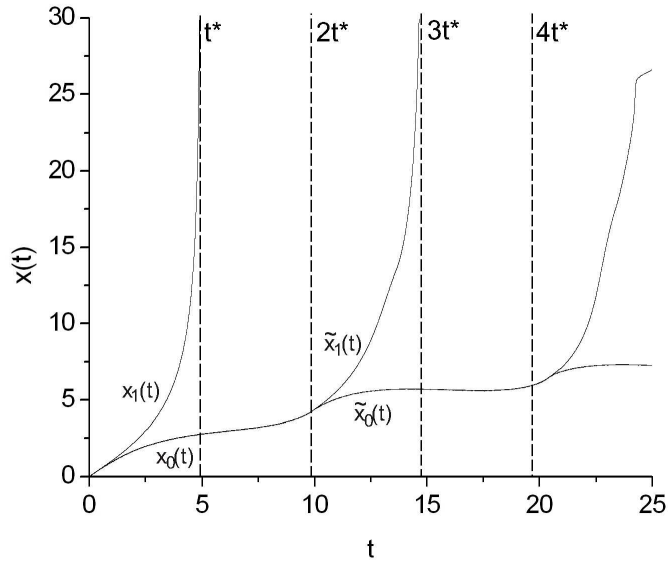}
\caption{Trajectories of zeros.}\label{duze czasy}
\end{center}
\end{figure}

Zero $x_0(t)$ decelerates up to time $t\approx t^*$, then it
accelerates up to $t\approx 2t^*$ when it reaches the velocity
$v=1$. At this moment a very interesting phenomenon occurs. Zero
$x_0(t)$ splits into a pair of zeros $\widetilde{x}_0(t)$ and
$\widetilde{x}_1(t)$ that move similarly to $x_0(t)$ and $x_1(t)$.
Numerical values of $\widetilde{x}_1(t)$ increase rapidly for
$t\rightarrow 3t^*-$, which suggests that function
$\widetilde{x}_1(t)$ has a vertical asymptote at $t=3t^*$. This
phenomenon seems to have an almost periodic character.

One can show by an elementary calculation (we skip the proof) that
the point $x_0$, at which the trivial solution ($f=0$) and a
nonzero solution are matched so that the first right-hand spatial
derivative of $f$ at $x_0$ is equal to zero, can not move with
velocity $v\geq 1$. The solution $f(x,t)$, that consists of the
partial solutions matched at the points that move with the
velocity $v\neq 1$, has to be smooth (i.e first derivative is
continuous) at the matching points. We can conclude that an
accelerating zero, that is a matching point of the trivial partial
solution $f=0$ and other nonzero partial solution, cannot move
faster than $v=1$ without changing its character. This change
means that the first spatial derivative of the field $f(x,t)$ is
nonzero at this point. It is possible provided that an additional
zero, that moves with velocity $v<1$, appears. The segment of $x$
axis between those two zeros widens. Such segment is a support of
a partial solution of the second kind. At the moment when $x_0$
reaches the velocity $v=1$ the solution $f(x,t)$ returns with good
approximation to its self-similar initial data (\ref{dane}) (see
Fig. \ref{odtworzenie}).
\begin{figure}[h!]
\begin{center}
\includegraphics[width=0.7\textwidth]{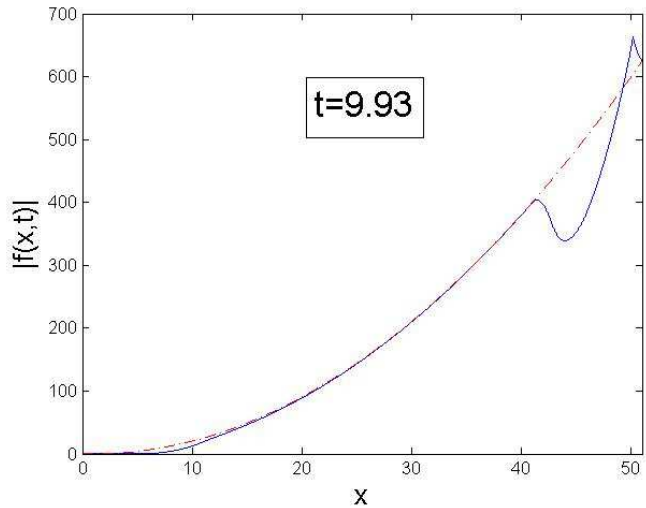}
\caption{Absolute value of the solution $f(x,t)$ at $t=2t^*$ - the
solid line, and the parabola of self-similar initial data - dashed
line. A discrepancy at the right-hand side is a pure numerical
effect caused by finite size of the grid.}\label{odtworzenie}
\end{center}
\end{figure}

\section{Summary}

A study of perturbed field-theoretic models with V-shaped
potential and the scaling symmetry can give some knowledge that
are properties of more general field models with potentials that
are sharp at its minima. One of the simplest example of such
models is the s-G model with perturbation that introduces an
additional linear term to the field equation, which gives the s-K-G
model. We have found that such a perturbation is responsible for
some new effects. For the same initial data the differences
between solutions in the s-G and the s-K-G models are caused by
this linear term.

The first observation comes from analysis of trajectories of
zeros. The zeros of the self-similar solutions in the s-G model
move with constant velocities, whereas velocity of zeros of
solutions in the s-K-G model is not monotonous. Moreover, their
accelerations and decelerations also depend on time. Nevertheless,
for small times $t\ll 1$ both the acceleration and the
deceleration of zeros are proportional to the perturbation
parameter $\lambda$. This fact can be useful as a phenomenological
criterion that allows us to calculate the parameter $\lambda$ from
experimental data.

We can also point out the second qualitative difference between
these models. There are partial solutions in the s-K-G model that
supports expand from zero to infinite size within a finite time.
It is possible provided that the other partial solutions disappear
simultaneously . For instance, the partial solution
$\phi_1=\frac{1}{4}(x^2-t^2)$ at $[t,\infty)$ in the s-G model
disappears when $t\rightarrow \infty$, whereas the partial
solution $f_+$ in the s-K-G model disappears when $t\rightarrow
\frac{\pi}{2\sigma}$.

One of the most interesting results that have been obtained from
approximated formulae for partial solutions $f_-^R(x,t)$ and
$f_-^L(x,t)$ is an observation that these partial solutions
contain singular terms proportional to $\lambda^{-1/3}$. It
suggests that (apart from technical obstacles) the partial
solution $f_-(x,t)$ can not be represented in the series form in
the similar way to $f_+(x,t)$.

The last observation is mainly numerical. The trajectories of
zeros of the solution $f(x,t)$ are almost periodic. We have
proposed here a hypothesis that the period has the value
$2t^*=\frac{\pi}{\sigma}$, where $t^*$ has been obtained from
analytical calculations - it is the characteristic time for which
the trajectory of zero $x_1(t)$ has a vertical asymptote. Our
hypothesis agrees quite well with the numerical data. The
periodicity for longer than investigated times is an open
question. In our study the solutions $f(x,0)$ and $f(x,2t^*)$ are
very similar which means that the solution $f(x,t)$ returns to the
self-similar initial data even though the scaling symmetry is
broken.

An important and open question is the behaviour of solutions of the
s-K-G model for other self-similar initial data or more general
initial data. In the group of general initial data the most
interesting are these for which an initial field configuration has
a finite energy (the energy of the self-similar solutions has an
infinite value). Finally, there are, of course, a variety of
perturbations of the potential $V(\phi)=|\phi|$ that can be
studied, nevertheless, it is clear that for most of them
analytical results can be obtained only in approximation.

\section{Acknowledgements}
The author is grateful to Tomasz Tyranowski for his assistance in
numerical work and valuable discussions as well as to Henryk
Arod\'z,  Andrzej Wereszczy\'nski and Jakub Lis for discussions
and remarks.

\end{document}